\documentclass{article}
\usepackage{times}
\def\ref{\noindent\hangafter=1\hangindent=1cm}
\input epsf
\begin{document}
\centerline{\large\bf Reanalysis of the OGLE-I Observations with the Image
Subtraction Method}
\centerline{\large\bf I. Galactic bar fields MM1-A, MM1-B, MM7-A, and
MM7-B\footnote{Based on observations obtained in the Las Campanas
Observatory of the Carnegie Institution of Washington.}}

\vspace{0.3cm}
\centerline{by}

\vspace{0.3cm}
\centerline{A.~P i g u l s k i,\quad Z.~K o {\l } a c z k o w s k i\quad
            and\quad G.~K o p a c k i}

\vspace{0.3cm}
\centerline{Wroc{\l }aw University Observatory, Kopernika 11, 51-622
            Wroc{\l }aw, Poland}
\centerline{E-mails: (pigulski, kolaczk, kopacki)@astro.uni.wroc.pl}

\vspace{0.5cm}
\centerline{\it Received ...}

\vspace{1cm}
\centerline{ABSTRACT}

\vspace{0.5cm}
{\small
As a result of the reanalysis of the OGLE-I observations by means of
the image subtraction method, we present the first part of a catalog,
consisting of the data for variable stars in four Galactic fields
observed by the OGLE-I, viz.~MM1-A, MM1-B, MM7-A, and MM7-B.  In
total, 2016 variable stars have been found.  This increased the number
of known variable in these fields stars more than twofold.  We comment
on the detectability of the variable stars in previous studies.

Some interesting findings are also discussed.  Among others, we found
45 $\delta$ Scuti stars (38 are new) including several multiperiodic
objects.  Detailed analysis of the light curves of 47 RR Lyrae stars
(24 are new detections) allowed us to find five stars which exhibit
nonradial pulsations and one RRd star.  Three RRab stars are members
of the Sagittarius dwarf galaxy (Sgr dSph).  We also find four objects
which are probably galactic RV Tauri stars and one W Virginis star which
seems to belong to Sgr dSph.  This is the first Population II Cepheid
found in this satellite galaxy and dSph other than the Fornax one.

For eclipsing EW-type binaries, which are the most abundant variables
in our catalog, we investigate the amplitude and period distributions.
A comparison with the previous OGLE-I catalogs indicates that we found
more stars with smaller amplitudes.  Finally, in addition to the two
microlensing events discovered previously in these fields, we find
five more.

\vspace{0.2cm}

}
%%%

\vspace{0.5cm}
\centerline{\bf 1. Introduction}

\vspace{0.5cm}
Thousands of variable stars have been discovered in the central
regions of our Galaxy by the Optical Gravitational Lensing Experiment
(OGLE), a long-term microlensing project which in June 2001 entered
its third stage of operation, OGLE-III\footnote{See {\tt
http://www.astrouw.edu.pl/\~{}ogle/}}.  In the first part of the
project (OGLE-I; Udalski {\it et al.}~1992), observations were
collected in four seasons (1992--1995) with the 1-m Swope telescope of
the Las Campanas Observatory.  Twenty 15$^\prime$ $\times$ 15$^\prime$
fields (see below) have been monitored frequently enough to allow
reliable time-series analysis.

During the second part of the OGLE project (OGLE-II; Udalski et
al.~1997b), carried out in the years 1997--2000, observations were
obtained with a dedicated 1.3-m Warsaw Telescope at the same
observatory.  A new, 2K $\times$ 2K CCD detector was used in a
drift-scan mode, resulting in a 2K $\times$ 8K frame covering a field
of view of 14$^\prime$ $\times$ 57$^\prime$.

Almost all Galactic fields observed by the OGLE are very rich in
stars.  The profile-fitting photometry in such fields is severely
affected by crowding.  This reduces detectability of variable stars.
Fortunately, a new method of finding variable stars in crowded fields,
based on image subtraction, has been developed when the OGLE-II
project was in operation (Alard and Lupton 1998; Alard 2000).  Since
the method proved to be very efficient in the detection of variable
stars in crowded fields, it was implemented by Wo\'zniak (2000) in his
Difference Image Analysis (DIA) package and then applied to the
OGLE-II observations of 49 Galactic fields.  This analysis resulted in
a catalog consisting of $\sim$200,000 candidates for variables in the
central regions of the Galaxy (Wo\'zniak {\it et al}.~2002).

The methods, direct profile fitting and image subtraction, are
complementary and, when used together, provide complete information on
the incidence of the variability in the observed fields.

Although most OGLE-I fields have been covered by the OGLE-II and are
observed within the OGLE-III projects carried out with better
detectors, there are still several reasons why the reanalysis of the
OGLE-I observations with the image subtraction can yield valuable
results.  Firstly, in view of the completness the method provides, its
application would certainly yield new variable stars.  Next, for those
stars which were already discovered, the DIA would provide better
photometry.  From the point of view of a study of the long-term
behavior of variable stars, it is a very good reason.  Finally, a
comparison of the detectability by the image subtraction and
profile-fitting methods is possible.

The OGLE-I data cover twenty fields (BWC, BW1 to BW11, MM1-A, MM1-B,
MM3, MM5-A, MM5-B, MM7-A, MM7-B, and GB1) with a number of frames
suitable for time-series analysis.  Except for the fields BW9--BW11,
GB1, MM1-A/B and MM3 which were not observed in 1992, observations in
four seasons, 1992--1995, are available.

In this series of papers we plan to provide a complete information on
the variability of stars in these twenty OGLE-I fields, based on the
results of reanalysis of the OGLE-I observations by means of the DIA.
We will include also the less numerous (20--50 per field) $V$-filter
observations, reported for variable stars by Szyma\'nski {\it et
al.}~(2001) only.  In the present paper, the first of the series, we
report the results of the search in four fields, MM1-A, MM1-B, MM7-A,
and MM7-B, located in the Galactic bar, a few degrees north-east of
the Baade's Window (BW).  The second paper will include fields MM3,
MM5-A, MM5-B and GB1, while the third one, all twelve BW fields.
Since the adjacent OGLE-I fields slightly overlap, we shall report the
observations of a given variable star including data from all
overlapping fields.

The variables discovered earlier in the OGLE-I data are briefly
summarized in the next section.  The method we use to search for
variable stars is outlined in Sect.~3, whereas the fourth section
describes the entries of the catalog we provide.  Section 5 includes
the comparison of the detectability of variable stars.  Finally,
comments on different types of variable stars are given in Sect.~6.

\vspace{0.5cm}
\centerline{\bf 2. Known Variables}

\vspace{0.5cm}
As far as the catalogs of variable stars are concerned, so-far
published results from the OGLE-I data are the following.  The most
important one is the Catalog of Periodic Variable Stars (CPVS)
containing 2861 periodic variables with the mean $I_{\rm C}$ magnitude
brighter than 18~mag.  The catalog has been published for sixteen of
the twenty fields mentioned above in five papers (Udalski {\it et
al}.~1994; 1995a,b; 1996; 1997a).  A similar catalog was not published
for four fields, MM1-A, MM1-B, GB1, and MM3.  Additions to the CPVS
include a list of 116 long-period and non-periodic variable stars
published by \.Zebru\'n (1998, hereafter Z98), 11 variables discovered
in the vicinity of 20 microlensing events found in the OGLE-I data
(Wo\'zniak and Szyma\'nski 1998, hereafter WS98), and recently
published catalog of 2084 periodic variables (Szyma\'nski {\it et
al.}~2001; hereafter SKU01), mainly fainter than $I_{\rm C} \approx$
18~mag.  Variable stars in the latter paper were divided into two
categories:  contact binaries and pulsating variable stars.

The known variable stars discovered in the four OGLE-I fields under
consideration (MM1-A, MM1-B, MM7-A, MM7-B) are summarized in Table 1.
For the OGLE-I microlensing events (MLE), their numbers from WS98
(preceded by ``\#'') are given.  Other variables found by WS98 are
labeled as VAR.  SKU01 discoveries are reported separately for their
contact binaries (EW) and pulsating stars (P).  Figure 1 shows the
equatorial coordinates of variable stars.  As we mentioned earlier,
CPVS includes variable stars for the fields MM7-A and MM7-B (Udalski
{\it et al}.~1997a, hereafter U97), but not for MM1.
%
%   === This is Figure 1 ===
%
\begin{figure}[t]
\epsfbox{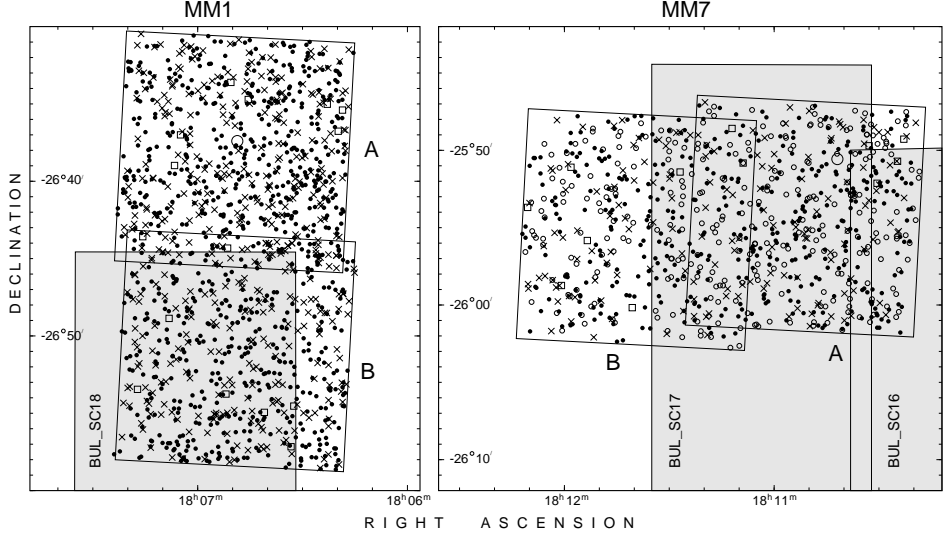}
\caption{Location of variable stars in fields MM1 and MM7.  The
symbols have the following meaning:  circles, stars in the CPVS (U97);
crosses, stars found by SKU01; squares, long-period variables
discovered by Z98 and WS98, large circles, the OGLE-I microlensing
events, OGLE\#9 in field MM7-A and OGLE\#14 in field MM1-A.  The new
variable stars we found are shown as small dots.  The grey areas
indicate the OGLE-II fields (labeled).}
\end{figure}

%
%     =========  Table 1 ==============
%
\begin{center}
{\small
T a b l e\quad 1\\

Variables discovered by U97, Z98, WS98 and SKU01 in four analyzed
OGLE-I fields:  MM1-A, MM1-B, MM7-A, and MM7-B.

\vspace{2mm}
\begin{tabular}{crrccr}
\hline\noalign{\smallskip}
OGLE-I& \multicolumn{4}{c}{Number of variables (only new)}
&  \\ \cline{2-5}\noalign{\smallskip} field & U97 & Z98 & WS98 (VAR +
MLE) & SKU01 (EW + P) & Total \\
\noalign{\smallskip}\hline\noalign{\smallskip}
MM1-A & --- &  9 & 0 + 1 (\#14) & 156 + 78 & 244 \\
MM1-B & --- &  5 & 1 + 0        & 131 + 55 & 192 \\
\noalign{\smallskip}\hline\noalign{\smallskip}
MM7-A & 182 &  4 & 0 + 1 (\#9)  &  72 + 15 & 274 \\
MM7-B & 112 &  7 & 0 + 0        &  48 + 24 & 191 \\
\noalign{\smallskip}\hline\noalign{\smallskip}
Total & 294 & 25 & 1 + 2        & 407 + 172 & 901 \\
\noalign{\smallskip}\hline
\end{tabular}

}
\end{center}

It has to be noted that rectangle areas around MM1 and MM7 OGLE-I
fields shown in Fig.~1 contain no objects listed in the General
Catalog of Variable Stars (GCVS, Kholopov {\it et al.}~1988), the Name
Lists 67--76 which followed that catalog (see Kazarovets {\it et
al.}~2001), and the New Catalogue of Suspected Variable Stars
(Kukarkin {\it et al.}~1982) with the Supplement (Kazarovets {\it et
al.}~1998).

The MM1 fields have been partly covered by OGLE-II field BUL\_SC18 and
MACHO field 101, whereas the MM7 fields were covered by OGLE-II fields
BUL\_SC16 and BUL\_SC17 and MACHO field 178 (see Fig.~1).

The third part of the All Sky Automated Survey (ASAS-3, Pojma\'nski
2002) will also cover this area, but the catalog of variables from
this part of the sky has not been published yet.  It will, however,
contain variable stars which are brighter than the saturation limit
for stars in the OGLE-I data ($I_{\rm C} \approx$ 14 mag).

\vfill\eject
%\vspace{0.5cm}
\centerline{\bf 3. The Method}

\vspace{0.5cm}
The calibrated OGLE-I observations used in our analysis have been
kindly made accessible to us by Prof.~Andrzej Udalski.  They were
obtained through two filters, matching Gunn $i$ and Johnson $V$ bands
(the former is easily transferable to Cousins $I_{\rm C}$, see Udalski
{\it et al.}~1992).  Most observations were carried out with the Gunn
$i$ filter.  Before starting reductions we excluded some frames
polluted by scattered light and those with short exposure times.  The
number of frames used in the final calculations is given in Table 2.
In this table, we specify the frames which were used to calculate the
reference image.  The frames used as reference ones in astrometric
interpolations are shown with boldface in the fiftth column.

Using the information provided with the frames, we have chosen frames
with the smallest seeing and low background to calculate the reference
image needed for image subtraction (See Table 2).  Typically, 6--7
frames in the Gunn $i$ filter and 3--4 frames in the Johnson $V$
filter were used for this purpose.

%
%     =========  Table 2 ==============
%
\begin{center}
{\small
T a b l e\quad 2\\

Data for the OGLE-I fields studied in this paper

\vspace{2mm}
\begin{tabular}{ccccrl}
\hline\noalign{\smallskip}
OGLE-I& \multicolumn{2}{c}{Coordinates (field center)} &&
Number & Frames used to create \\
field & $\alpha_{\rm 2000}$ & $\delta_{\rm 2000}$ & Filter &
of frames & a reference image\\
\noalign{\smallskip}\hline\noalign{\smallskip}
MM1-A & 18$^{\rm h}$06$^{\rm m}$49.5$^{\rm s}$&
$-$26$^{\rm o}$37$^\prime$59$^{\prime\prime}$ & $V$ & 23 &
10295, 10609, 11470 \\
&&&$i$ &143& 10053, 10108, {\bf 10460}, 10610, \\
&&&&& 10865, 11469, 11779\\
\noalign{\smallskip}\hline\noalign{\smallskip}
MM1-B & 18$^{\rm h}$06$^{\rm m}$49.7$^{\rm s}$ &
$-$26$^{\rm o}$50$^\prime$57$^{\prime\prime}$ & $V$ & 22 & 10210,
10297, 11471 \\
&&&$i$ &132& 10054, {\bf 10109}, 10250, 10822, \\
&&&&& 10866, 11290, 11468\\
\noalign{\smallskip}\hline\noalign{\smallskip}
MM7-A & 18$^{\rm h}$10$^{\rm m}$53.2$^{\rm s}$ &
$-$25$^{\rm o}$54$^\prime$15$^{\prime\prime}$ & $V$ & 26 & 11263,
11606, 12016, 12211 \\
&&&$i$ &168& 10204, 10443, {\bf 10547}, 11456, \\
&&&&& 11590, 11734\\
\noalign{\smallskip}\hline\noalign{\smallskip}
MM7-B & 18$^{\rm h}$11$^{\rm m}$39.8$^{\rm s}$ &
$-$25$^{\rm o}$55$^\prime$05$^{\prime\prime}$ & $V$ & 26 & 11264,
11303, 12017, 12212 \\
&&&$i$ &171& 10101, 10158, {\bf 10444}, 10548, \\
&&&&& 11455, 11637, 11735\\
\noalign{\smallskip}\hline
\end{tabular}

}
\end{center}

The procedure of the image subtraction has been explained in detail in
the original papers (Alard and Lupton 1998, Alard 2000, Wo\'zniak
2000) and will not be repeated here.  We only mention that the image
subtraction algorithm requires a reference image which is subsequently
convolved with a kernel in order to match seeing in a given image and
then subtracted from it.  As a result, difference image is obtained.
The difference images can be stacked to form a ``variability map''
which is used for detecting variables.  The reference image and the
variability map for a given field (and filter) were obtained by means
of the DIA package of Wo\'zniak (2000).  The 2K $\times$ 2K images
were divided into 8 $\times$ 8 = 64 square sub-fields which have been
processed independently.  However, the variable search procedure
implemented in the DIA turned out not to detect many apparent
variable stars in the variability map.  We therefore replaced it by
the Daophot (Stetson 1987) FIND routine run on the variability map.
The output was a list of candidates for variable stars which has been
sent back to the DIA which calculated light curve for each candidate.

In an ideal case, all stars appearing in the variability map should be
variable.  For several reasons this was not the case.  For the OGLE-I
data, a huge number of spurious candidates for variables has been
produced by the chip defects.  Typically, about 20,000 variable
candidates per field have been obtained.  Only 2--3\% of them appeared
to be true variables.  In order to extract them in an automatic way,
we calculated the AoV periodogram of Schwarzenberg-Czerny (1996) in
the range between 0 and 30~d$^{-1}$ for each candidate.  In order to
increase the detection efficiency, the light curves have been first
cleaned for outliers (we used 5$\sigma$ clipping) and observations
with large errors (they were not rejected from further analysis,
however).  The AoV periodogram provides the parameter $\theta$ which
has been used as a threshold.  It has to be noted that this procedure
allows also detection of non-periodic variables since such variables
yield high peaks in the vicinity of 0 and 1~d$^{-1}$.

After this step, we were usually left with 700--1000 candidates for
variables in a field.  A light curve of each of these stars was
checked visually, period(s) refined, and the classifications made.
The script we used created a single plot showing:  (i) flux difference
as a function of time, (ii) phase diagram (the correct period could be
selected from the list of highest peaks in periodogram, allowing also
its doubling, a common case for eclipsing binaries), (iii) a part of
the periodogram in the vicinity of the strongest peak, (iv) a part of
the reference image and variability map with the analyzed star
centered.  At this stage, observations in both filters were used.

\vspace{0.5cm}
\centerline{\bf 4. The Catalog}

\vspace{0.5cm}
In total, we found 2016 variable stars in the four OGLE-I fields
analyzed in this paper (see Table 3).  As we shall discuss in Sect.~5,
we have recovered 84\% of variable stars discovered by previous
investigators and, in addition, found 1255 new variables.  This means
that we increased the number of variables in fields MM1 and MM7 more
than twofold.
%
%     =========  Table 3 ==============
%

\begin{center}
{\small
T a b l e\quad 3\\

The number of variables discovered in fields MM1 and MM7 according to
class.  The thirteen classes of variables are defined in Sect.~4.2.

}{\tiny
\vspace{2mm}
\begin{tabular}{rrrrrrrrrrrrrrr}
\hline\noalign{\smallskip}
& \multicolumn{13}{c}{T y p e \quad o f \quad v a r i a b i l i t
y}&\\ \cline{2-14} \noalign{\smallskip}
Field & DSCT & RRc & RRab & CEP & Mira & EW & EB & EA & SIN & NSIN &
PER & APER & MLE & Total \\
\noalign{\smallskip}\hline\noalign{\smallskip}
&\multicolumn{14}{c}{K n o w n}\\
\noalign{\smallskip}\hline\noalign{\smallskip}
MM1 &  2 &  5 &--- &--- &--- & 289 & 18 & 12 & --- & --- & 15 & 4 & 1
& 346 \\
MM7 &  5 &  5 & 13 & 1  &--- & 220 & 20 & 62 &  6 & 7 &  68 & 8 & ---
& 415\\
\noalign{\smallskip}\hline\noalign{\smallskip}
Both&  7 & 10 & 13 & 1  &--- & 509 & 38 & 74 &  6 & 7 & 83 & 12& 1 &
761 \\
\noalign{\smallskip}\hline\noalign{\smallskip}
&\multicolumn{14}{c}{N e w}\\
\noalign{\smallskip}\hline\noalign{\smallskip}
MM1 & 26 &  6 & 17 & 4 & 9 & 229 & 23 & 122 & 45 & 29 & 290 & 43 & 3 &
846 \\
MM7 & 12 & ---&  1 & --- & 2 & 151 & 6 & 39 & 31 & 25 & 119 & 21 & 2 &
409 \\
\noalign{\smallskip}\hline\noalign{\smallskip}
Both& 38 &  6 & 18 &  4 & 11 & 380 & 29 & 161 & 76 & 54 & 409 & 64 & 5
& 1255\\
\noalign{\smallskip}\hline\noalign{\smallskip}
&\multicolumn{14}{c}{A l l}\\
\noalign{\smallskip}\hline\noalign{\smallskip}
MM1 & 28 & 11 & 17 & 4 & 9 & 518 & 41 & 134 & 45 & 29 & 305 & 47 & 4 &
1192 \\
MM7 & 17 &  5 & 14 & 1 & 2 & 371 & 26 & 101 & 37 & 32 & 187 & 29 & 2
& 824 \\
\noalign{\smallskip}\hline\noalign{\smallskip}
Both& 45 & 16 & 31 & 5 & 11 & 889 & 67 & 235 & 82 & 61 & 492 & 76 & 6
& 2016\\
\noalign{\smallskip}\hline
\end{tabular}

}
\end{center}

The two MM1 fields, MM1-A and MM1-B, slightly overlap (Fig.~1).  The
same is true for the two MM7 fields.  This means that some variables
can be found in both fields.  We therefore report the variables using
designations which indicate only a combined field, MM1 or MM7.

The parameters of all variable stars we found are given in Tables 4a
(field MM1) and 4b (field MM7), available in the full form in the {\it
Acta Astronomica Archive} (see cover page) and our WWW page
\begin{quote}
{\tt http://www.astro.uni.wroc.pl/ldb/ogle1/}
\end{quote}
Here, only a sample page of Table 4b is shown.  The tables include:
star name, its equatorial coordinates, classification (see Sect.~4.2),
period(s), reference magnitude and color, number of data points and
cross-identifications with previous OGLE catalogs.  If no
cross-identification is given, the variable is a new one.

We also present our results in a graphical form.  A sample plot is
displayed in Fig.~2.  The plots showing light curves and finding
charts can be found on our WWW page as gzipped PostScript files.  The
following file names have been used:  $<${\it field}$>$\_$<${\it
va\-ria\-bility class}$>$[\_$<${\it sequential number}$>$].ps; for
example file ``MM1\_EW\_2.ps'' includes the second page with the light
curves and finding charts for EW-type eclipsing binaries in the field
MM1.  If there was only one page for a given class, no sequential
number was given.

%
%     =========  Table 4b ==============
%
\begin{center}

T a b l e\quad 4b\\

Variable stars in field MM7.  The full table is available in an
electronic form only.  Here, the entries for the first 50 stars are
shown as an example.

{\tiny

\vspace{2mm}
\begin{tabular}{cccllrrccl}
\hline\noalign{\smallskip}
Var.~name & $\alpha_{2000.0}$ & $\delta_{2000.0}$ & Var.~type & Period
[d] & $N_I$ & $N_V$ & $I_{\rm C}$ &
($V-I_{\rm C}$)&  Cross-identifications \\
\noalign{\smallskip}\hline\noalign{\smallskip}
MM7-0001 & 18 10 34.39  &  $-$25 48 28.5  &  DSCT          & 0.04604239 & 165 & 25 & 15.526 &  0.969 &                                               \\
MM7-0002 & 18 10 26.50  &  $-$25 48 36.8  &  DSCT          & 0.04959108 & 157 &  0 & 17.718 &  1.563 &                                               \\
MM7-0003 & 18 11 47.40  &  $-$25 48 08.8  &  DSCT          & 0.05584052 & 164 & 19 & 16.026 &  1.383 &                                               \\
         &              &               &                & 0.05701629 &     &    &        &        &                                               \\
         &              &               &                & 0.0636008  &     &    &        &        &                                               \\
MM7-0004 & 18 11 04.87  &  $-$25 56 47.6  &  DSCT          & 0.06580976 & 162 & 26 & 14.979 &  0.793 &                                               \\
MM7-0005 & 18 11 55.37  &  $-$25 55 19.5  &  DSCT HADS     & 0.06583299 & 167 & 23 & 17.445 &  1.417 & OGLE MM7-B V87 (DSCT)                         \\
MM7-0006 & 18 10 20.77  &  $-$25 52 07.8  &  DSCT          & 0.06674510 & 165 & 26 & 16.985 &  1.161 &                                               \\
         &              &               &                & 0.0686684  &     &    &        &        &                                               \\
MM7-0007 & 18 10 22.39  &  $-$25 48 02.6  &  DSCT          & 0.06808241 & 159 & 25 & 16.331 &  1.176 &                                               \\
MM7-0008 & 18 12 02.43  &  $-$25 53 32.2  &  DSCT          & 0.0732052  & 160 & 22 & 15.761 &  1.190 &                                               \\
MM7-0009 & 18 10 28.65  &  $-$25 49 11.3  &  DSCT HADS     & 0.07884756 & 166 & 25 & 17.242 &  1.143 & OGLE MM7-A V103 (DSCT)                        \\
MM7-0010 & 18 10 41.83  &  $-$25 54 56.0  &  DSCT          & 0.08860681 & 165 & 26 & 16.813 &  1.150 &                                               \\
         &              &               &                & 0.0904652  &     &    &        &        &                                               \\
MM7-0011 & 18 10 49.84  &  $-$25 56 47.4  &  DSCT          & 0.0934265  & 160 & 26 & 15.616 &  1.097 &                                               \\
MM7-0012 & 18 11 08.12  &  $-$26 00 56.3  &  DSCT          & 0.09588007 & 166 & 25 & 16.379 &  1.224 &                                               \\
         &              &               &                & 0.1247319  &     &    &        &        &                                               \\
         &              &               &                & 0.0766860  &     &    &        &        &                                               \\
MM7-0013 & 18 10 48.67  &  $-$25 52 58.5  &  DSCT HADS     & 0.15885904 & 150 & 21 & 16.932 &  1.396 & OGLE MM7-A V88 (DSCT)                         \\
MM7-0014 & 18 12 06.28  &  $-$25 52 13.1  &  DSCT          & 0.1677467  & 169 & 23 & 17.135 &  1.793 &                                               \\
MM7-0015 & 18 10 32.14  &  $-$25 59 55.7  &  DSCT          & 0.1723239  & 159 & 26 & 15.465 &  2.060 &                                               \\
MM7-0016 & 18 10 22.80  &  $-$25 53 04.7  &  DSCT HADS/EW  & 0.1972421  & 162 &  0 & 18.882 &   ---  & SKU MM7A.19894 (EWs)                          \\
MM7-0017 & 18 10 26.01  &  $-$25 58 54.7  &  DSCT/EW       & 0.2066995  & 155 & 25 & 17.433 &  1.757 & OGLE MM7-A V139 (EW) =\\
&&&&&&&&& = SKU MM7A.4811 (EWs) \\
\noalign{\smallskip}\hline\noalign{\smallskip}
MM7-0018 & 18 10 54.50  &  $-$25 49 42.4  &  RRc           & 0.22759215 & 147 & 26 & 15.942 &  1.125 & OGLE MM7-A V44 (RRc)                          \\
MM7-0019 & 18 10 49.72  &  $-$25 48 24.0  &  RRc           & 0.2749935  & 161 & 26 & 17.048 &  1.568 & OGLE MM7-A V94 (RRc)                          \\
MM7-0020 & 18 11 33.45  &  $-$25 55 08.3  &  RRc           & 0.2867497  & 157 & 21 & 15.802 &  1.330 & OGLE MM7-B V14 (RRc)                          \\
MM7-0021 & 18 10 35.61  &  $-$25 52 51.8  &  RRc           & 0.2878246  & 156 & 26 & 16.180 &  1.534 & OGLE MM7-A V47 (RRc)                          \\
MM7-0022 & 18 10 39.85  &  $-$25 59 15.0  &  RRc NR        & 0.3066806
& 168 & 26 & 16.030 &  1.251 & SKU MM7A.34501 (EWa,\\
&&&&&&&&& P = 0.64918 d)           \\
         &              &               &                & 0.3245838  &     &    &        &        &                                               \\
\noalign{\smallskip}\hline\noalign{\smallskip}
MM7-0023 & 18 10 48.96  &  $-$25 56 29.8  &  RRab          & 0.4436107  & 146 & 26 & 16.349 &  1.175 & OGLE MM7-A V26 (RRab)                         \\
MM7-0024 & 18 10 32.07  &  $-$25 59 36.2  &  RRab          & 0.4938028  & 168 & 26 & 15.619 &  1.741 & OGLE MM7-A V25 (RRab)                         \\
MM7-0025 & 18 11 18.64  &  $-$26 00 41.3  &  RRab          & 0.52984105 & 330 & 48 & 16.068 &  1.449 & OGLE MM7-A V20 (RRab)                         \\
MM7-0026 & 18 11 12.65  &  $-$25 49 37.5  &  RRab          & 0.5304776  & 318 & 49 & 15.969 &  1.593 & OGLE MM7-A V34 (RRab)                         \\
MM7-0027 & 18 11 40.09  &  $-$25 51 02.8  &  RRab          & 0.5332197  & 164 & 23 & 15.860 &  1.319 & OGLE MM7-B V13 (RRab)                         \\
MM7-0028 & 18 12 03.44  &  $-$25 58 22.2  &  RRab          & 0.5391315  & 165 & 23 & 15.569 &  1.492 & OGLE MM7-B V6 (RRab)                          \\
MM7-0029 & 18 11 00.07  &  $-$25 57 38.0  &  RRab          & 0.5408001  & 155 & 26 & 15.025 &  1.534 & OGLE MM7-A V2 (RRab)                          \\
MM7-0030 & 18 11 30.96  &  $-$25 56 01.8  &  RRab          & 0.5538256  & 170 & 23 & 16.350 &  1.263 &                                               \\
MM7-0031 & 18 12 00.30  &  $-$25 53 19.9  &  RRab          & 0.5593891  & 153 & 22 & 15.629 &  1.423 & OGLE MM7-B V8 (RRab)                          \\
MM7-0032 & 18 11 14.73  &  $-$25 51 46.6  &  RRab          & 0.5670927  & 162 & 23 & 15.847 &  1.295 & OGLE MM7-A V16 (RRab)                         \\
MM7-0033 & 18 11 51.14  &  $-$25 58 11.8  &  RRab          & 0.590028   & 167 & 23 & 15.793 &  1.209 & OGLE MM7-B V7 (RRab)                          \\
MM7-0034 & 18 11 44.86  &  $-$25 58 03.1  &  RRab          & 0.616314   & 165 &  0 & 18.545 &  2.197 & SKU MM7B.75698 (P)                            \\
MM7-0035 & 18 11 47.90  &  $-$25 52 23.2  &  RRab          & 0.6180556  & 160 & 23 & 17.369 &  1.649 & OGLE MM7-B V63 (RRab)                         \\
MM7-0036 & 18 11 36.71  &  $-$25 49 06.7  &  RRab          & 0.8812044  & 167 & 23 & 15.607 &  1.622 & OGLE MM7-B V12 (RRab)                         \\
\noalign{\smallskip}\hline\noalign{\smallskip}
MM7-0037 & 18 11 12.01  &  $-$25 55 51.1  &  EW            & 0.23476895 & 320 & 48 & 18.086 &  1.859 & SKU MM7A.168828 (EWs)                         \\
MM7-0038 & 18 11 05.38  &  $-$25 55 03.1  &  EW            & 0.2634260  & 157 & 25 & 14.841 &  1.673 &                                               \\
MM7-0039 & 18 11 25.78  &  $-$26 00 21.4  &  EW            & 0.2637791  &  96 &  0 &   ---  &   ---  &                                               \\
MM7-0040 & 18 11 50.72  &  $-$25 52 39.1  &  EW            & 0.2670858  & 167 & 23 & 18.591 &  2.089 & SKU MM7B.108382 (EWa)                         \\
MM7-0041 & 18 10 43.62  &  $-$26 00 44.2  &  EW            & 0.2690071  & 160 &  0 &   ---  &   ---  & SKU MM7A.65167 (EWa)                          \\
MM7-0042 & 18 10 54.51  &  $-$25 49 11.2  &  EW            & 0.2722634  & 162 &  0 &   ---  &   ---  &                                               \\
MM7-0043 & 18 11 22.51  &  $-$25 53 10.7  &  EW            & 0.2779473  & 138 &  0 & 17.318 &  1.749 &                                               \\
MM7-0044 & 18 10 49.20  &  $-$25 51 12.7  &  EW            & 0.2788253  & 155 &  0 & 18.230 &   ---  &                                               \\
MM7-0045 & 18 11 35.76  &  $-$25 58 39.6  &  EW            & 0.2794520  & 167 &  0 & 18.183 &  2.087 &                                               \\
MM7-0046 & 18 10 54.08  &  $-$25 54 38.3  &  EW            & 0.2842176  & 163 & 24 & 18.689 &  1.613 & SKU MM7A.109242 (P)                           \\
MM7-0047 & 18 11 35.90  &  $-$25 54 59.1  &  EW            & 0.2877905  & 159 & 22 & 17.576 &  1.783 &                                               \\
MM7-0048 & 18 11 09.85  &  $-$25 55 39.6  &  EW            & 0.2884005
& 326 & 49 & 17.538 &  1.652 & OGLE MM7-A V138 (EW)\\
&&&&&&&&& = SKU MM7A.168261 (EWa)  \\
MM7-0049 & 18 11 51.36  &  $-$26 00 56.4  &  EW            & 0.2931527  & 157 &  0 & 19.493 &  2.063 &                                               \\
MM7-0050 & 18 10 58.47  &  $-$25 51 26.3  &  EW            & 0.2932039  & 154 &  0 & 19.623 &   ---  &                                               \\
\noalign{\smallskip}\hline
\end{tabular}

}
\end{center}

Typically, three panels for a given star are presented in a plot
(see Fig.~2).  For periodic stars and stars with a dominating period
shorter than 25 days the left-hand panel shows the light curve as a
phase diagram.  For periodic stars with dominating period longer than
25 days and stars classified as APER, MLE and MIRA the light curve is
presented as a function of heliocentric Julian day (this is not shown
in Fig.~2).  The star name (Sect.~4.1), period (P) or dominating
period (D) as well as cross-identifications are also plotted in this
panel.  The right-hand top panel shows a finding chart of 99 $\times$
99 pixels (43$^{\prime\prime}$ $\times$ 43$^{\prime\prime}$) from the
frame used as a reference one and oriented roughly in such a way that
north is up, east to the left.  The variable star is centered and
encircled with a light-grey circle.  Some additional informations on
the star (classification, coordinates, magnitudes) are given in the
right-hand bottom panel.

For some stars, more than one plot is presented.  This has been done
for short-period multiperiodic stars, when in addition to a phase
diagram showing original data folded with the main period, we also
show all detected periodicities separately (that is, folded after
removing the other ones, see Fig.~2 for MM7-0022).  In a few cases,
when a strong trend was apparent, we removed it and show the phase
diagram once again, but freed from the trend.  In such cases plots
contain a remark ``trend removed''.  When the data from two
overlapping fields are combined, an encircled letter ``C'' is plotted
right of our star name (see plots for MM7-0025 and MM7-0037 in
Fig.~2).
%
%   === This is Figure 2 ===
%
\begin{figure}[h]
\epsfbox{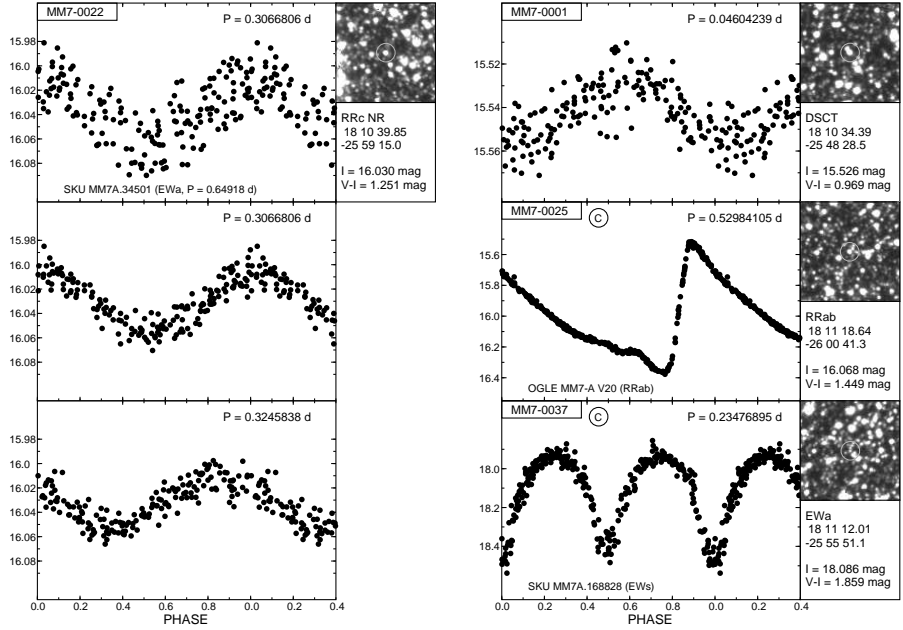}
\caption{A sample page of the graphical data which include $I_{\rm
C}$-filter light curves and finding charts for variables in our
catalog.  Typically, data for 20 or 30 stars are plotted in a single
page.  Here, for clarity, we show only the data for four selected
stars in the field MM7.  The ordinate is the $I_{\rm C}$-filter
magnitude; phase 0.0 was chosen arbitrarily.  See text for more
explanations.}

\end{figure}

There were some variable stars which were too faint or too crowded to
be detected by Daophot in the reference image.  In such cases, a
correct transformation from fluxes to magnitudes (see Sect.~4.4) could
not be performed.  For these stars we do not give the mean magnitudes
and/or colors in plots and tables.  However, in order to show the
light curves of all stars on the magnitude scale, we assumed that the
magnitudes of these stars are close to the limiting magnitude in the
reference frame and did the transformation.  This helps to estimate
the amplitude, but it must be remembered that the magnitudes of these
stars could be unreliable.

\vspace{0.5cm}
\indent {\it 4.1. Designation of variables}

\vspace{0.5cm}
The already known variables from the OGLE catalogs will be referred to
using designations introduced by their authors.  The CPVS introduced
the names of variable stars consisting of a field designation, letter
``V'' and a sequential number.  In this catalog (see, e.g., U97) the
stars were arranged according to the decreasing mean brightness.  The
microlensing events were numbered separately (OGLE \#1 to OGLE \#20;
WS98).  Z98 introduced new designations for his long-period variables
(e.g., OGLE-LT-V011).  On the other hand, SKU01 and WS98 use for their
variables the numbers from the OGLE photometric database.  We shall
refer to these variables using the following abbreviations:  SKU for
SKU01 and WS for WS98, original field name and star number, e.g.,
SKU\,MM1B.24085.  The OGLE catalogs provide the coordinates and
finding charts, so that the proper identification or
cross-identification is rather easy.

The extensions of some of the above-mentioned numbering systems could
be somewhat misleading.  Therefore, we decided to introduce our own
designations consisting of the field name (without division into part
A or B, if two adjacent fields were observed) and a four-digit number,
e.g., MM1-0019.  The stars in Tables 4a and 4b and in the plots
described above are presented in groups, according to the
classification scheme explained in the next subsection.  Within a
given group, the stars are arranged according to the increasing period
(or dominating period), except for aperiodic stars and microlensing
events which are presented in the order of increasing right ascension.

\vspace{0.5cm}
\indent {\it 4.2. The classification scheme}

\vspace{0.5cm}
Many variable stars reported in this paper are periodic or
multiperiodic.  In the process of classification of these stars we
have taken into account the following features:  period(s), amplitude,
the shape of the light curve, and---in some cases---the location in
the color-magnitude diagram.  As far as the stars with periods below
$\sim$1 d are concerned, the classification was quite easy and rather
unambiguous.  Among stars with longer periods, only the Miras are
easily recognizable.

The following types of variability have been attributed to the
periodic pulsating stars:
\begin{itemize}
\item DSCT -- $\delta$ Scuti stars.  If the light curve had the
$I_{\rm C}$-filter full amplitude larger than 0.1~mag, the designation
HADS (high-amplitude $\delta$ Scuti star) was given in addition to
DSCT.  Note that some of the HADS, especially with the shortest
periods, can be Population II pulsators, known as SX Phoenicis stars.
\item RRc -- RR Lyrae stars of Bailey type c.  We also included
one RRd star (double-mode RR Lyrae star) into this group.  For RRc
stars with nonradial modes present, the ``NR'' designation has been
added (this applies also to RRab stars).
\item RRab -- RR Lyrae stars of Bailey type ab.
\item CEP -- Cepheid variables and RV Tauri stars.
\item MIRA -- Miras.
\end{itemize}

Classifying eclipsing binaries we have adopted an old system based on
the morphology of the light curve which divides these stars into three
categories.  These are:  W Ursae Majoris-type (EW), $\beta$~Lyrae-type
(EB), and Algol-type (EA).  However, the distinction between EB and EA
or EB and EW depends on the quality of the light curve.  Consequently,
the classification of the eclipsing binaries is somewhat arbitrary for
some stars.  If a distinct O'Connell effect was observed, the
designation ``OC'' was added.

The presence of large scatter causes that two stars with light curves
of exactly the same shape can be classified in a different way.  There
is, however, no way to avoid this problem.  Hence, for some stars
double classification is given (e.g.~EW/DSCT) in the order of the
authors' preference.

The classification of stars that show periodicities longer that
$\sim$1~d is more difficult.  Most of these stars were classified as
miscellaneous (MISC) in the CPVS.  Some are strictly periodic, but
there are too many possible types of variability in a given range of
periods to make a correct classification.  If the light curve is
non-sinusoidal or there is an evidence of multiperiodicity, a final
decision can be sometimes made.  Such is the case of some ellipsoidal
variables.  We finally decided to divide the stars into the
three following categories:
\begin{itemize}
\item SIN -- Stars with sinusoidal light curves showing the evidence
of a single periodicity.  This category may include such variables as
the monoperiodic Slowly Pulsating B-type (SPB) stars, $\alpha^2$~Canum
Venaticorum stars, $\gamma$~Doradus stars, also some ellipsoidal
variables (with half the orbital period) and chromospherically active
stars, etc.
\item NSIN -- Stars with strictly periodic light curves, which are
evidently non-sinusoidal.  The ellipsoidal variables (``ELL''
designation added) can present such light curves, but other types of
variability are also possible in this group.
\item PER -- Stars with a dominating periodicity, but showing also
amplitude and/or phase changes or the periodic changes superimposed on
the variability on a longer time-scale.  A majority of semi-regular
variables fall into this category.
\end{itemize}
In addition, stars showing aperiodic behaviour were classified as:
\begin{itemize}
\item APER -- Stars that show aperiodic light variations.  It has to
be noted that stars with linear light changes that might arise from
high proper motion (see Eyer and Wo\'zniak 2001, Soszy\'nski {\it et
al.}~2002), but also from the intrinsic light variation, were not
included in our catalog.
\item MLE -- Microlensing events.
\end{itemize}

\vspace{0.5cm}
\indent {\it 4.3. Multiperiodicity}

\vspace{0.5cm}
The correct period (or periods) of a light variation was decided
during the classification process and was found by means of the AoV
periodogram using solely the $I_{\rm C}$-filter data.  The final
period reported in Tables 4a and 4b represents, however, a refined
value obtained from a nonlinear least-squares solution.  The
least-squares fit was done also for non-sinusoidal light curves; in
such a case the appropriate number of harmonics was fitted to the
data.  Depending on the accuracy of the period determination, periods
are given with a different number of digits.

The search for multiperiodicity was made only for some groups of
stars, namely DSCT, RRc, RRab, CEP and some SIN and PER stars with the
shortest periods.  This reflects the scientific interests of the
authors.  The results are discussed in Sect.~6.  Thus, it is still
possible that some stars we do not announce to be multiperiodic,
especially from the SIN and PER groups have, in fact, multiple
periods.

\vspace{0.5cm}
\indent {\it 4.4. Conversion from the fluxes to magnitudes}

\vspace{0.5cm}
The DIA provides differential fluxes in arbitrary units.  In order to
convert the fluxes into magnitudes we first derived the instrumental
magnitudes of all stars in the reference frame by means of a standard
profile-fitting photometry, namely using Daophot package of Stetson
(1987).  This was done for each of the four fields (MM1-A, MM1-B,
MM7-A, MM7-B) and the two filters (Johnson $V$, Gunn $i$) separately.
The instrumental magnitudes, $p=\{v, i\}$, were used to derive the
coefficients of the transformation equation:
\begin{equation}
P_k = -2.5\log (f_0 + f_k) + C,
\end{equation}
where $f_0 = 10^{-0.4(p + C_2)}$ was the reference flux expressed in
the same units as the differential fluxes, $f_k$, for the $k$th
measurement.  $P_k$ stands for the corresponding $k$th measurement
taken from the $P = \{V, I_{\rm C}\}$ photometry of U97 and SKU01.
Using stars in common with U97 and/or SKU01, we first derived the
values of $f_0$ and $C$ for each star.  It appeared that $C$ varies
slightly with the instrumental magnitude, $p$.  Therefore, we adopted
a linear form of the dependence, $C = a + b\cdot(p - 15.0)$, and
derived the coefficients $a$ and $b$ by means of the least squares.
Then, with $C(p)$ fixed, the $f_0$ values were recalculated and
finally, the constant $C_2$ was determined.  The final coefficients
are given in Table 5.  There are many possible sources of the weak
dependence of $C$ on $p$, including small nonlinearities of the
detector indicated by Udalski {\it et al.}~(1992).  We shall not
speculate on that.  Taking into account the small color terms in the
transformation to the standard $VI_{\rm C}$ system (Udalski {\it et
al.}~1992), our transformation should give reliable standard
magnitudes as well.

\vfill\eject
%
%     =========  Table 5 ==============
%
\begin{center}
{\small
T a b l e\quad 5\\

The coefficients of the transformation from differential fluxes to the
standard $VI_{\rm C}$ magnitudes for fields MM1 and MM7.

\vspace{2mm}
\begin{tabular}{cccccc}
\hline\noalign{\smallskip}
Field & Filter & $a$ [mag] & $b$ & $C_2$ [mag] & RSD [mag]\\
\noalign{\smallskip}\hline\noalign{\smallskip}
MM1A & $V$ & 25.7827 $\pm$ 0.0188 & 0.0395 $\pm$ 0.0136 & $-$22.8063
$\pm$ 0.0167 & 0.096 \\
& $I_{\rm C}$ & 25.0856 $\pm$ 0.0087 & 0.0238 $\pm$ 0.0094 &
$-$22.7038 $\pm$ 0.0148 & 0.074 \\
\noalign{\smallskip}\hline\noalign{\smallskip}
MM1B & $V$ & 25.7786 $\pm$ 0.0361 & 0.0668 $\pm$ 0.0218 & $-$22.8267
$\pm$ 0.0170 & 0.130 \\
& $I_{\rm C}$ & 25.0458 $\pm$ 0.0103 & 0.0225 $\pm$ 0.0098 &
$-$22.7954 $\pm$ 0.0159 & 0.079 \\
\noalign{\smallskip}\hline\noalign{\smallskip}
MM7A & $V$ & 25.4173 $\pm$ 0.0154 & 0.0465 $\pm$ 0.0124 & $-$22.7987
$\pm$ 0.0193 & 0.025 \\
& $I_{\rm C}$ & 25.0083 $\pm$ 0.0066 & 0.0392 $\pm$ 0.0075 &
$-$22.7745 $\pm$ 0.0068 & 0.072 \\
\noalign{\smallskip}\hline\noalign{\smallskip}
MM7B & $V$ & 25.5806 $\pm$ 0.0274 & 0.0646 $\pm$ 0.0217 & $-$22.6753
$\pm$ 0.0240 & 0.062 \\
& $I_{\rm C}$ & 25.0345 $\pm$ 0.0054 & 0.0145 $\pm$ 0.0059 &
$-$22.7809 $\pm$ 0.0081 & 0.048 \\
\noalign{\smallskip}\hline
\end{tabular}

}
\end{center}

Having derived the coefficients $a$, $b$, and $C_2$ for each
field/filter combination we could transform the differential fluxes
for all variable stars into the standard $VI_{\rm C}$ magnitudes.
However, since the instrumental magnitudes were derived by means of
the profile-fitting photometry in a crowded field from a {\it single}
reference frame, the errors of the transformation could be quite
large.  Thus, the magnitudes should be used with caution.  The
photometry (both the differential fluxes and the transformed
magnitudes) is available on the web page given above.

The reference magnitudes of variable stars were also transformed to
the standard $VI_{\rm C}$ magnitudes.  They are given in Tables 4a and
4b and in our plots (see Fig.~2).  Very few variable stars were not
found in the reference frame.  In these cases we assumed they have the
instrumental magnitude close to the limiting magnitude of the
photometry in a given reference frame.  For these stars the reference
magnitudes and/or colors were not included in Tables 4a and 4b.  We
used them only for plotting a star's light curve, so that in our plots
all stars would have the ordinate given in magnitudes.

\vspace{0.5cm}
\indent {\it 4.5. Equatorial coordinates}

\vspace{0.5cm}
Equatorial coordinates of variable stars were derived using their
positions in the variability map.  For each field we identified about
4000 stars from the Hubble {\it Guide Star Catalog}, ver.~2.2.1, and
used a third-order polynomial to transform each coordinate into an
appropriate equatorial one.  The accuracy of the final position
depends primarily on the accuracy of the position in the variability
map; we estimate it to be of the order of 0.2$^{\prime\prime}$.  There
are systematic differences up to 1$^{\prime\prime}$ between our
coordinates and those given in OGLE catalogs.  They occur probably
because of using different astrometric catalogs as a reference.

\vspace{0.5cm}
\centerline{\bf 5. Comments on Detectability}

\vspace{0.5cm}
It is interesting to check how many variable stars presented in the
OGLE catalogs mentioned in Sect.~2 could be recovered by means of the
DIA.  Out of the 294 variables from the CPVS, detected in the two MM7
fields (Table 1), we have recovered 284, that is about 97\%.  The
fact that the remaining 10 stars were not detected can be explained as
follows:
\begin{itemize}
\item[(i)] The DIA provides no photometry for stars
which are overexposed in the reference frame.  This is the reason why
the two brightest stars in the CPVS, MM7-A V1 and MM7-B V1, were
missed.
\item[(ii)] Since some areas around overexposed pixels are marked as
unusable, stars which fall there will have no DIA photometry either.
This was the case for MM7-A V112 and MM7-B V100.
\item[(iii)] There are some shifts between the frames.  The DIA
software detects only these variables which are located in the region
common with the reference frame.  Three CPVS stars, MM7-A V8, V56 and
V114, were outside this region.
\item[(iv)] If the star has a low amplitude and/or is faint, it may be
too faint in the variability map to be detected by the search
procedure.  This was the case MM7-A V172.
\item[(v)] This rather rare case happens when Daophot search
procedure cannot resolve variables which are very close to each other.
This was the case for MM7-A V133 which is very close to MM7-A V77.  In
consequence, the former star was not detected.
\end{itemize}

In addition, we found one CPVS star (MM7-A V86) which can be barely
seen in the variability map but was detected.  However, after careful
analysis we concluded that it shows no clear evidence for variability.
We therefore regard it as constant so it does not appear in our
catalog.

Z98 reports 26 variable stars in fields MM1 and MM7 (Table 1).  One,
OGLE-LT-V106, is identical with MM7-A V116, so that 25 stars are new
detections.  We recover them all but OGLE-LT-V108 because of point
(ii).  In addition, two other stars, OGLE-LT-V001 and OGLE-LT-V024,
are probably not real variables, but rather high-proper motion stars,
showing nearly linear change of brightness if constant position is
used to get the magnitudes (Eyer and Wo\'zniak 2001; Soszy\'nski {\it
et al.}~2002).  We do not include them either.

We also detect a variable found by WS98 in field MM1-B (WS
MM1B.198992).  Of the two microlensing events, OGLE\#9 in field MM7-A
and OGLE\#14 in MM1-A, we detect only the latter.  The former was too
faint in the variability map to be detected (see point (iv) above).

SKU01 report 591 new variable stars in fields MM1 and MM7, but some
have double designations, so that there are actually 579 (407
classified as eclipsing, EW, and 172 as pulsating, P; see Table 1)
new variables.  We recovered 100 out of 120 EW stars (83\%) and 20 out
of 39 P stars (51\%) in the field MM7.  In fields MM1 the detection
efficiencies are slightly higher.  This is because the CPVS did not
include this field, so the bright variable stars located there were
discovered by SKU01.  We found 250 out of 287 EW stars (87\%) and 83
out of 133 P stars (62\%) in field MM1.  The main reason for
non-detections is point (iv) above:  stars are too faint in the
variability map to be detected.  Some other stars are not detected
because they are very close to overexposed stars (point (ii)).

While the small difference between the detection efficiency in fields
MM1 (79\%) and MM7 (75\%) is understandable in view of the comments
given above, the large difference in the detectability of the EW
(86\%) and P stars (60\%) is rather a surprise.  Both classes have
similar distribution in brightness, so that the brightness difference
is not the explanation.  Moreover, out of 103 P stars we recovered in
this study, only eight (8\%) were classified as pulsating stars in our
catalog, RRc or DSCT.  Remaining P variables are almost exclusively
EW-type eclipsing binaries.  Recalling the fact that among variables
classified by SKU01 as EW we found four pulsating stars, we conclude
that the algorithm used by SKU01 to distinguish eclipsing and
pulsating stars by means of the Fourier coefficients apparently did
not work well.  The stars classified as P have probably lower
amplitudes than the EW stars, as can be seen from Fig.~14 of SKU01.
This explains both the large spread of Fourier coefficients in the
decomposition shown by SKU01 (their Fig.~1) and much lower detection
efficiency of P stars in comparison with EW stars in our catalog.

\vspace{0.5cm}
\centerline{\bf 6. Comments on Individual Variable Stars}

\vspace{0.5cm}
A thorough analysis of the contents of the presented catalog is out of
scope of this paper.  However, taking into account our personal
interests, we shall discuss in detail the results for some of the
variables we found.

\vfill\eject
\indent {\it 6.1. $\delta$ Scuti stars}

\vspace{0.3cm}
Some $\delta$~Scuti stars were included in the CPVS, but because this
search was limited to frequencies up to only 10~d$^{-1}$ and stars
brighter than $I_{\rm C} \approx$ 18~mag, it is clear that mainly
High-Amplitude $\delta$~Scuti (HADS) stars had been detected and that
many low-amplitude $\delta$ Scuti stars could be still found.  This is
indeed the case:  we found 45 $\delta$~Scuti stars, of which only 7
were known from U97 and SKU01.  Seven stars show clearly multiperiodic
behaviour; as many as three significant periodicities were found in
MM7-0003 and MM7-0012.  Fifteen stars in field MM1 and four stars in
MM7 could be classified as the HADS.  The location of $\delta$ Scuti
stars in the colour-magnitude diagram (CMD) is shown in Fig.~3.  While
the low-amplitude variables are mostly disk stars, the HADS, lying 1
to 3~mag below RR Lyraes, seem to be mainly at the bulge distance.
The same was found by Alcock {\it et al.}~(2000a) from their analysis
of a much larger sample of the bulge HADS.  If they are normal,
evolved main-sequence HADS, this would be an indication of the
presence of a not so old (age $<$ 1.5--2 Gyr) stellar population at
those distances.  However, it is more likely they are SX Phoenicis
stars of the old disk/bulge population or a mixture of both.
%
%   === This is Figure 3 ===
%
\begin{figure}[t]
\epsfbox{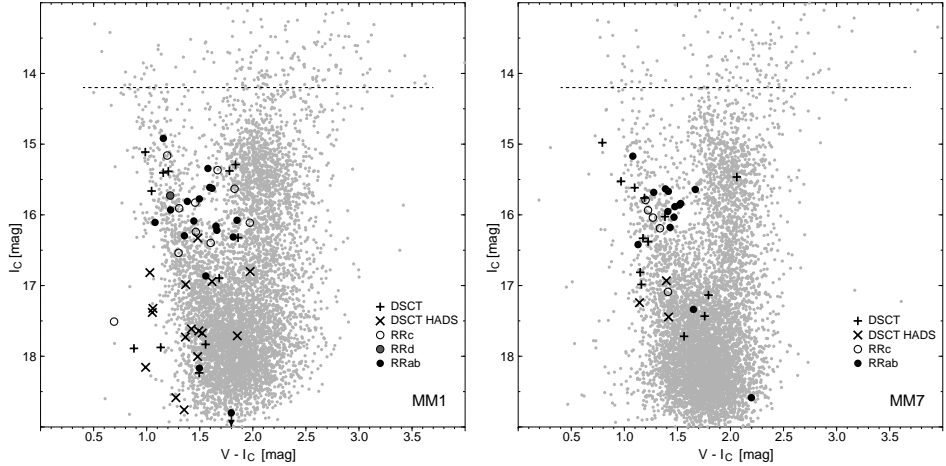}
\caption{Color-magnitude diagrams for field MM1 (left) and MM7
(right).  Only about 10\% of stars with the best photometry
($\sigma_{\rm V-I} <$ 0.1~mag) is shown (as grey dots).  Positions of
$\delta$~Scuti and RR Lyrae stars are plotted with different symbols.
The saturation limit is shown as a dashed line.}
\end{figure}

\vspace{0.3cm}
\indent {\it 6.2. RR Lyrae stars}

\vspace{0.3cm}
These stars have large amplitudes.  Therefore, we expected that, at
least in fields MM7, no new variables would be detected.  It appeared,
however, that even in these fields we were able to find a new RRab
star, MM7-0030.  It avoided earlier detection because of its location
close to several bright stars.  Out of the remaining 5 RRc and 13 RRab
stars in fields MM7, CPVS reported sixteen and SKU01 added two.  One
of the latter, MM7-0022 = SKU MM7A.34501, is a very interesting
example of an RRc star showing two close modes with periods of $P_1$ =
0.3066806~d and $P_2$ = 0.3245838~d and almost equal amplitudes in
$I_{\rm C}$ (see the left side of Fig.~2).  It was classified by SKU01
as an EWs variable (EW-type eclipsing binary with similar depth of
both eclipses) and period equal to 2$P_2$.

In fields MM1 we found 28 RR Lyrae variables.  Only five of them (all
of type RRc) were reported by SKU01.  This is because SKU01 focused on
the short-period EW-type binaries and not only RRab, but also other
variables with longer periods were not presented there (see Table 3).
The sample of RR Lyrae variables in field MM1 includes 11 RRc and 17
RRab stars.  Among them there are five stars showing multiperiodic
behaviour.  The periods of all multiperiodic RR Lyrae stars we found
are listed in Table 5.  This table is therefore supplementary to the
recent work of Moskalik and Poretti (2003).  These authors analyzed
the whole CPVS sample of 215 RR Lyrae stars finding 38 multiperiodic
stars, including a double-mode RRd star, BW7\,V30.  We also find one
RRd star, MM1-0033, with the radial first-overtone to fundamental
period ratio equal to 0.737571 $\pm$ 0.000005.

Another very interesting RR Lyrae star is MM1-0046, an RRab type star
showing three close equally spaced frequencies, characteristic for the
Blazhko effect.  In this respect the star is very similar to the star
BW6\,V20 analyzed in detail by Moskalik and Poretti (2003).

%
%     =========  Table 5 ==============
%
\begin{center}
{\small
T a b l e\quad 5\\

Multiperiodic RR Lyrae variables in fields MM1 and MM7.  In
parentheses, the r.m.s.~errors of the last two digits of preceding
numbers are given.

\vspace{2mm}
\begin{tabular}{ccccc}
\hline\noalign{\smallskip}
Name in the & &Primary period & Secondary period & Frequency
difference\\
catalog & Type$^{\rm a}$ & $P_0 = f_0^{-1}$ [d] & $P_1 =
f_1^{-1}$ [d] & $\Delta f = f_1-f_0$ [d$^{-1}$]  \\
\noalign{\smallskip}\hline\noalign{\smallskip}
MM1-0033 & RRd & 0.3049400(08) & 0.413438(07) & $-$0.860593(16) \\
MM1-0034 & RR1-$\nu$1 & 0.3063829(05) & 0.306245(04) & $+$0.001470(14)
\\
MM1-0044 & RR0-$\nu$1 & 0.5179836(15) & 0.521618(14) & $-$0.013451(26)
\\ MM1-0046 & RR0-BL & 0.5336114(12) & 0.536120(07) & $-$0.008769(13)
\\          &        &           & 0.531090(08) & $+$0.008897(16) \\
MM1-0051 & RR0-$\nu$1 & 0.5816272(14) & 0.579520(21) & $+$0.006249(36)
\\ \noalign{\smallskip}\hline\noalign{\smallskip}
MM7-0022 & RR1-$\nu$1 & 0.3066806(14) & 0.3245838(16) & $-$0.179853(7)
\\ \noalign{\smallskip}\hline\noalign{\smallskip}
\end{tabular}

$^{\rm a}$ According to Alcock {\it et al.}'s (2000c)
classification scheme.}

\end{center}
The position of the RR Lyrae variables in the CMD is shown in Fig.~3.
A majority of them is located at the bulge distance as indicated by
previous investigators (Udalski 1998, Alcock {\it et al.}~1998, Morgan
{\it et al.}~1998), but three (MM1-0047, MM1-0054, MM7-0034) that are
$\sim$2~mag fainter than the rest, are members of the Sagittarius
dwarf galaxy (Sgr dSph, Mateo {\it et al.}~1995; Alard 1996; Alcock
{\it et al.}~1997, 1998).  The other three RR Lyrae stars lie in
between, they have 17.0 $< I_{\rm C} <$ 17.5~mag.  These are MM1-0031
= SKU MM1A.5623, MM7-0019 = OGLE\,MM7-A V94 and MM7-0035 =
OGLE\,MM7-B~V63.  The latter was indicated as a member of Sgr dSph by
Olech (1997).  However, their membership is rather doubtful and a
larger interstellar extinction could also explain their position in
the CMD.

\vspace{0.3cm}
\indent {\it 6.3. Eclipsing binaries}

\vspace{0.3cm}
Eclipsing binaries constitute about 60\% of stars in our catalog, of
them the EW-type stars are most common.  This enables us to make some
statistical considerations related to this type of stars.  The
properties of the OGLE-I EW-type binaries have been already studied by
Ruci\'nski (1995, 1997a,b, 1998a,b) using a sample of CPVS EW-type
systems from nine BW fields.  He found that these stars are almost
evenly distributed along the line of sight and only very few had
estimated distances comparable to that of the Galactic bulge.  This
conclusion is in good agreement with position of these stars in the
CMD (Fig.~4), where they populate the left branch consisting of the
main-sequence stars and stars close to the turn-off point at different
distances.  Several EW stars lie close to the red clump because of
blending with red stars.

Many other interesting results have been given in the above-mentioned
papers of Ru\-ci\'n\-ski with a clear indication that EW-type binaries
can play an important role in the study of stellar populations towards
and even in the bulge.  In his studies, the period and amplitude
distributions were used frequently to constrain properties of EW-type
stars.  It is, therefore, interesting to compare how the newer
discoveries in the OGLE-I fields, namely those of SKU01 and the one
reported in this paper, affect these two distributions.  Although in
this study we do not analyze BW fields, but fields in the Galactic
bar, it was shown by SKU01 (their Fig.~9) that at least the period
distribution of EW stars for the bar fields (MM1, MM3, MM5 and MM7)
does not differ significantly from that in the BW fields.

\vfill\eject
%
%   === This is Figure 4 ===
%
\begin{figure}[h]
\epsfbox{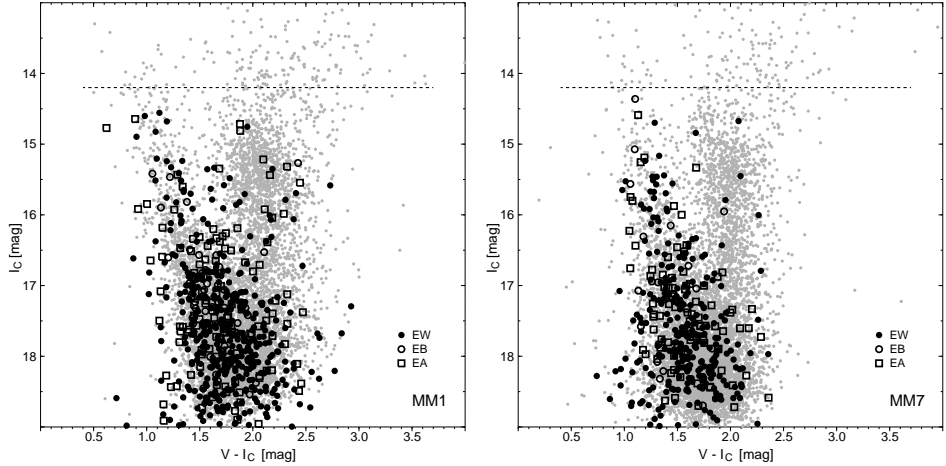}
\caption{The same as in Fig.~3, but with positions of eclipsing
binaries (EW, EB, EA) indicated.}
\end{figure}
%
%   === This is Figure 5 ===
%
\begin{figure}[h]
\epsfbox{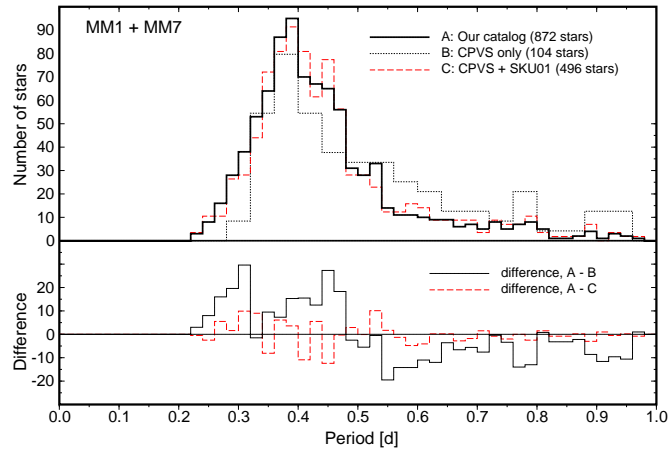}
\caption{A comparison of period distributions for EW-type binaries
with $P <$ 1~d in fields MM1 and MM7.  Three distributions are shown
in the top panel:  for 104 CPVS stars we found in MM7-A/B (B, dotted
line), for all known stars from CPVS and SKU01 we recovered (C, dashed
line) and finally, for all EW stars in our catalog (A, thick solid
line).  All distributions were normalized to the same number of stars.
In the lower panel, the differences between the distributions are
shown.}
\end{figure}

A comparison of the period distributions is shown in Fig.~5.  The
distribution calculated for EW stars with periods shorter than 1~d
from our catalog (solid line, 872 stars) was compared with two others:
that consisting of the CPVS EW stars from fields MM7 only (104 stars)
and all 496 already known stars (CPVS + SKU01).  All distributions
were normalized to the same number of stars.  As can be seen from the
figure, there is no significant difference between the distribution
for our and SKU01 EW binaries.  On the other hand, in comparison with
CPVS we have an excess of stars with $P <$ 0.5~d (and consequently a
deficit of stars with $P >$ 0.5~d).  Two conclusions can be drawn from
these differences:  (i) in comparison with profile-fitting photometry,
the DIA provides better completness, but does not reach fainter stars,
(ii) the difference in limiting magnitudes of the CPVS and SKU01/our
catalog resulted in different period distributions.  The latter can be
understood in view of the period-luminosity relation for the EW stars:
at a given distance, a contact binary with shorter orbital period
consists of smaller (and cooler) components and thus is fainter.
%
%   === This is Figure 6 ===
%
\begin{figure}[h]
\epsfbox{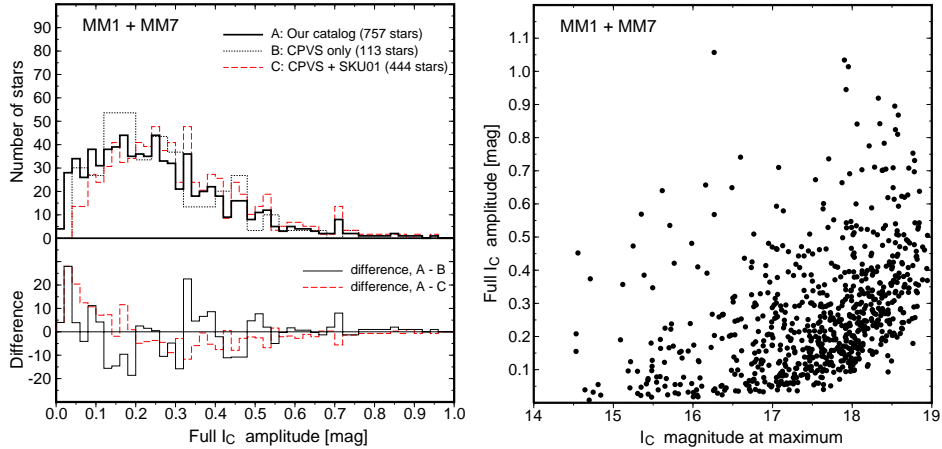}
\caption{Left:  A comparison of $I_{\rm C}$ amplitude distributions
for EW-type binaries with $I_{\rm C} <$ 19~mag.  See Fig.~5 for the
explanation of the data sets.  All distributions were normalized to
the same number of stars.  Right: Full $I_{\rm C}$-filter amplitudes
plotted as a function of the $I_{\rm C}$ magnitude at the phase of
maximum light.}
\end{figure}
%
%   === This is Figure 7 ===
%
\begin{figure}[h]
\epsfbox{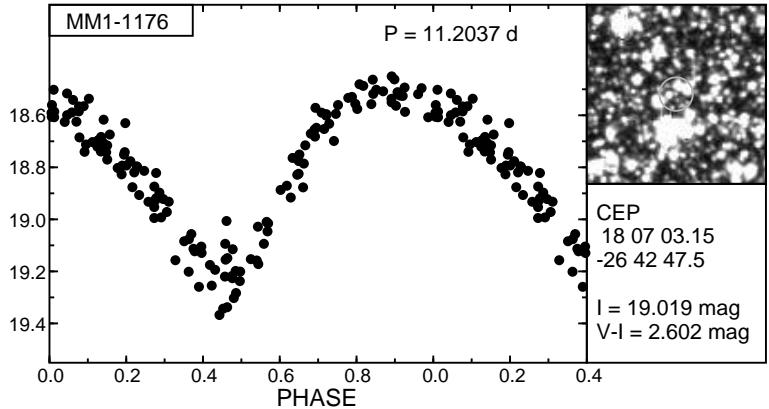}
\caption{The $I_{\rm C}$-filter light curve of MM1-1176, the first
W~Virginis star which is a likely member of the Sgr dSph.}
\end{figure}

We also compared the amplitude distributions (Fig.~6).  The analyzed
samples of stars are slightly different, because now we used all EW
stars with $I_{\rm C} <$ 19~mag at the phase of maximum light.  In
addition, we excluded all variable stars that were not detected in the
reference frame.  As can be seen from Fig.~6, we found relatively more
low-amplitude stars than did SKU01, although---as expected---many
faint EW stars with small amplitudes are not detected (see the right
panel of Fig.~6).

\vspace{0.3cm}
\indent {\it 6.4. Cepheids, RV Tauri Stars and Miras}

\vspace{0.3cm}
Classical Cepheids trace relatively young population, so they are not
expected to be present in the Galactic bulge.  Moreover, owing to
their high luminosities, they should be overexposed in the OGLE-I
observations even at the bulge distance.  However, some Population II
Cepheids and RV Tauri stars could be present and detected.  The latter
stars are not always easily recognized from long-period Population II
Cepheids and are closely related to them (Wallerstein 2002), so we
discuss them jointly.

There are four stars in our catalog which were classified as RV Tauri
stars (we use designation CEP for them as explained in Sect.~4.2).
All show alternating deeper and shallower minima in the light curve,
characteristic for this group (see, e.g., Pollard {\it et al.}~1996).
Their (doubled) periods range from about 23 days for MM7-0820 to 105~d
for MM1-1179.  The fifth star which was included in the CEP group,
MM1-1176, is the most interesting.  According to its period equal to
11.2~d, large amplitude and an asymmetric light curve (see Fig.~7), it
could be Population II Cepheid of W~Virginis type.  However, it is
very faint and red; $I_{\rm C} \approx$ 19.0~mag, $(V-I_{\rm C})
\approx$ 2.6~mag.  Assuming it has an intrinsic color $(V-I_{\rm C})_0
\approx$ 0.5~mag and adopting $A_{I_{\rm C}}$ = 1.57$E(V-I_{\rm C})$
(Dean {\it et al.}~1978) we get $I_{\rm C,0} \approx$ 15.7~mag.
Giving the mean distance modulus of Sgr dSph of 17--17.5~mag and its
extended structure, MM1-1176 could be indeed a W~Virginis-type star,
the first known in this galaxy.  So far, Population II Cepheids were
found only in one of the nine Milky Way dwarf spheroidal galaxies,
namely the Fornax one (Bersier and Wood 2002).  Since Population II
Cepheids are in a very short AGB or post-AGB evolutionary phase, they
can be found only in populous dwarf galaxies.  Fornax and Sagittarius
dSphs are the two most luminous galaxies of this type among the Milky
Way Galaxy satellites (van den Bergh 2000).

We find also 11 Miras, all are new variable stars.  They were,
however, detected mainly at the phase close to the minimum light.
%
%   === This is Figure 8 ===
%
\begin{figure}[h]
\epsfbox{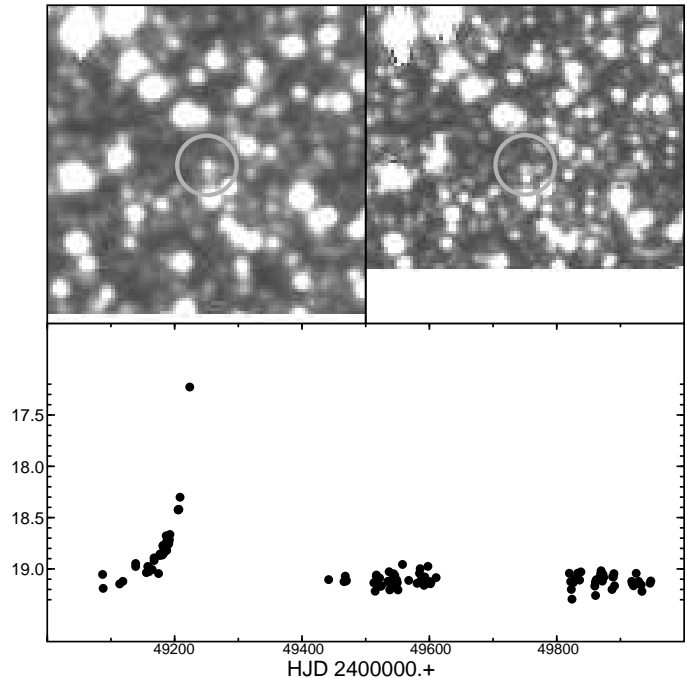}
\caption{The $I_{\rm C}$ light-curve of the microlensing event
MM1-1189.  The top left panel shows a part of the mr5262 frame made on
August 24, 1993 (the epoch of the brightest point in the light curve).
The top right panel shows the same area in the reference image which
was combined from the frames taken in 1995, about 2 years after the
event.}
\end{figure}

\vspace{0.3cm}
\indent {\it 6.5. Microlensing events}

\vspace{0.3cm}
Six stars showing a single event of brightness increase were found.
We cannot exclude the possibility that they are eruptive stars, but
the most plausible explanation is that they all are microlensing
events.  Only one of them, OGLE \#14, was known earlier (see WS98).
As an example, we show in Fig.~8 the light curve of MM1-1189, and also
two maps of the region, a reference frame made of the frames well
outside the epoch of the event and the frame when the star was
brightest.  The star is practically not visible in the reference
frame, being blended with another faint star.  It is, however, clearly
seen in the other frame despite much poorer seeing.

It is difficult to say how many new microlensing events will be found
in the remaining 16 OGLE-I fields, but because there were only two
known in fields MM1 and MM7, we expect at least to double their number
in the OGLE-I data.  The full analysis of these events will be carried
out as soon as all the OGLE-I fields are catalogued.  It is
interesting to note that the application of image subtraction led to
doubling the number of detected microlensing events also in OGLE-II
(Udalski {\it et al.}~2000, Wo\'zniak {\it et al.}~2001) and MACHO
(Alcock {\it et al.}~2000b) data.

\vspace{0.3cm}
\indent {\it 6.6. Other stars}

\vspace{0.3cm}
The largest number of new discoveries came for stars we classified as
SIN, NSIN, PER, and APER --- 85\% of them are new variables.
As can be seen in Fig.~9, they populate mostly the red branch of the
CMD.  However, some are relatively blue.  It is therefore possible
that some SIN or PER stars with periods shorter than 4~d, shown
as circles in Fig.~9, are slowly pulsating B-type (SPB) stars.

%
%   === This is Figure 9 ===
%
\begin{figure}[h]
\epsfbox{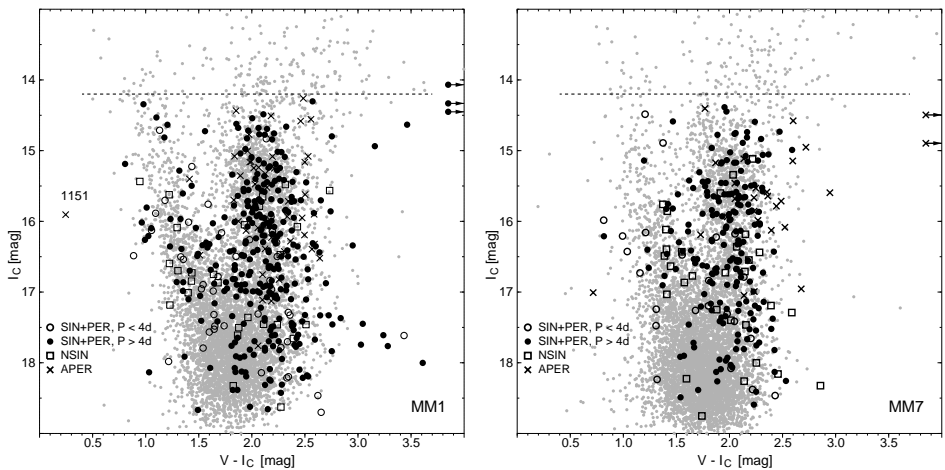}
\caption{The same as in Fig.~3, but with SIN, NSIN, PER, and APER
stars plotted.  The leftmost cross in the left panel, labeled
``1151'', denotes the position of the R Coronae Borealis variable
MM1-1151 = OGLE-LT-V035 (Z98).}
\end{figure}

There are two interesting stars among those classified as APER.  The
first one is MM1-1151 = OGLE-LT-V035 classified by \.Zebru\'n (1998)
as an R Coronae Borealis-type variable.  It is also the bluest
variable star we found (Fig.~9).  The other, MM7-0813, shows several
eruptions during the time covered by observations.  It is identical
with OGLE II BUL\_SC17\_4122, classified recently as a dwarf nova by
Cieslinski {\it et al}.~(2003).

\vspace{0.5cm}
\centerline{\bf Acknowledgements}

\vspace{0.3cm}
It is a great pleasure to express our appreciation to the whole OGLE
team for their superb work and especially for their policy to present
the data as soon as possible to the astronomical community.  In
particular, we thank Prof.~A.~Udalski for making available the
original OGLE-I observations of the Galactic fields.  We also thank
Dr.~P.~Wo\'zniak for allowing us to use his DIA package and
K.~\.Zebru\'n and I.~Soszy\'nski for explaining its details.  The
comments of Prof.~M.\,Jerzykiewicz were very helpful for improving the
paper.

This work was supported by the KBN grant 2\,P03D\,006\,19.

\vspace{0.5cm}
\centerline{REFERENCES:}

\vspace{0.3cm}
{\small
\ref Alard, C.~1996, {\it Astrophys.~J.}, {\bf 458}, L17.\par
\ref Alard, C.~2000, {\it Astron.~Astrophys.~Suppl.}, {\bf 144},
     363.\par
\ref Alard, C., and Lupton, R.H.~1998, {\it Astrophys.~J.}, {\bf
     503}, 325.\par
\ref Alcock, C.~{\it et al.}~1997, {\it Astrophys.~J.}, {\bf 474},
     217.\par
\ref Alcock, C.~{\it et al.}~1998, {\it Astrophys.~J.}, {\bf 492},
      190.\par
\ref Alcock, C.~{\it et al.}~2000a, {\it Astrophys.~J.}, {\bf 536},
     798.\par
\ref Alcock, C.~{\it et al.}~2000b, {\it Astrophys.~J.}, {\bf 541},
     734.\par
\ref Alcock, C.~{\it et al.}~2000c, {\it Astrophys.~J.}, {\bf 542},
     257.\par
\ref Bersier, D., and Wood, P.R.~2002, {\it Astron.~J.}, {\bf 123},
     840.\par
\ref Cieslinski, D., Diaz, M.P., Mennickent, R.E., and Pietrzy\'nski,
     G.~2003, {\it Publ.~Astron.~Soc.~Pacific}, {\bf 115}, 193.\par
\ref Dean, J.F., Warren, P.R., and Cousins, A.W.J.~1978, {\it Monhly
     Not.~Royal Astron.~Soc.}, {\bf 183}, 569.\par
\ref Eyer, L., and Wo\'zniak, P.R.~2001, {\it Monthly Not.~Royal
     Astron.~Soc.}, {\bf 327}, 601.\par
\ref Kazarovets, E.V., Samus, N.N., and Durlevich, O.V.~1998, {\it
     Inf.~Bull.~Var.~Stars} {\bf 4655}.\par
\ref Kazarovets, E.V., Samus, N.N., and Durlevich, O.V.~2001, {\it
     Inf.~Bull.~Var.~Stars} {\bf 5135}.\par
\ref Kholopov, P.N.~{\it et al.}~1988, {\it
     General Catalog of Variable Stars}, ed.~4, Moscow: Nauka
     Publishing House.\par
\ref Kukarkin, B.V.~{\it et al.}~1982, {\it New Catalogue of Suspected
     Variable Stars}, Moscow: Nauka Publishing House.\par
\ref Mateo, M., Kubiak, M., Szyma\'nski, M., Ka{\l }u\.zny, J.,
     Krzemi\'nski, W., and Udalski, A.~1995, {\it Astron.~J.}, {\bf
     110}, 1141.\par
\ref Morgan, S.M., Simet, M., and Bargenquast, S.~1998, {\it Acta
     Astron.}, {\bf 48}, 341.\par
\ref Moskalik, P., and Poretti, E.~2003, {\it Astron.~Astrophys.},
     {\bf 398}, 213.\par
\ref Olech, A.~1997, {\it Acta Astron.}, {\bf 47}, 183.\par
\ref Pojma\'nski, G.~2002, {\it Acta Astron.}, {\bf 52}, 397.\par
\ref Pollard, K.R., Cottrell, P.L., Kilmartin, P.M., and Gilmore,
     A.C.~1996, {\it Monthly Not.~Royal Astron.~Soc.}, {\bf 279},
     949.\par
\ref Ruci\'nski, S.M.~1995, {\it Astrophys.~J.}, {\bf 446}, L19.\par
\ref Ruci\'nski, S.M.~1997a, {\it Astron.~J.}, {\bf 113}, 407.\par
\ref Ruci\'nski, S.M.~1997b, {\it Astron.~J.}, {\bf 113}, 1112.\par
\ref Ruci\'nski, S.M.~1998a, {\it Astron.~J.}, {\bf 115}, 1135.\par
\ref Ruci\'nski, S.M.~1998b, {\it Astron.~J.}, {\bf 116}, 2998.\par
\ref Schwarzenberg-Czerny, A.~1996, {\it Astrophys.~J.}, {\bf
     460}, L107.\par
\ref Soszy\'nski, I.~{\it et al.}~2002, {\it Acta Astron.}, {\bf 52},
     143.\par
\ref Stetson, P.B.~1987, {\it Publ.~Astron.~Soc.~Pacific}, {\bf 99},
     191.\par
\ref Szyma\'nski, M., Kubiak, M., and Udalski, A.~2001, {\it Acta
     Astron.}, {\bf 51}, 259 (SKU01).\par
\ref Udalski, A.~1998, {\it Acta Astron.}, {\bf 48}, 113.\par
\ref Udalski, A., Szyma\'nski, M., Ka{\l }u\.zny, J., Kubiak, M., and
     Mateo, M.~1992, {\it Acta Astron.}, {\bf 42}, 253.\par
\ref Udalski, A., Kubiak, M., Szyma\'nski, M., Ka{\l }u\.zny, J.,
     Mateo, M., and Krzemi\'nski, W.~1994, {\it Acta Astron.}, {\bf
     44}, 317.\par
\ref Udalski, A., Szyma\'nski, M., Ka{\l }u\.zny, J., Kubiak, M.,
     Mateo, M., and Krzemi\'nski, W.~1995a, {\it Acta Astron.}, {\bf
     45}, 1.\par
\ref Udalski, A., Olech, A., Szyma\'nski, M., Ka{\l }u\.zny, J.,
     Kubiak, M., Mateo, M., and Krzemi\'nski, W.~1995b, {\it Acta
     Astron.}, {\bf 45}, 433.\par
\ref Udalski, A., Olech, A., Szyma\'nski, M., Ka{\l }u\.zny, J.,
     Kubiak, M., Mateo, M., Krzemi\'nski, W., and Stanek, K.Z.~1996,
     {\it Acta Astron.}, {\bf 46}, 51.\par
\ref Udalski, A., Olech, A., Szyma\'nski, M., Ka{\l }u\.zny, J.,
     Kubiak, M., Mateo, M., Krzemi\'nski, W., and Stanek, K.Z.~1997a,
     {\it Acta Astron.}, {\bf 47}, 1 (U97).\par
\ref Udalski, A., Kubiak, M., and Szyma\'nski, M.~1997b, {\it Acta
     Astron.}, {\bf 47}, 319.\par
\ref Udalski, A., \.Zebru\'n, K., Szyma\'nski, M., Kubiak, M.,
     Pietrzy\'nski, G., Soszy\'nski, I., and Wo\'zniak, P.R.~2000,
     {\it Acta Astron.}, {\bf 50}, 1.\par
\ref van den Bergh, S.~2000, {\it Publ.~Astron.~Soc.~Pacific}, {\bf
     112}, 529.\par
\ref Wallerstein, G.~2002, {\it Publ.~Astron.~Soc.~Pacific}, {\bf
     114}, 689.\par
\ref Wo\'zniak, P.R.~2000, {\it Acta Astron.}, {\bf 50}, 451.\par
\ref Wo\'zniak, P.R., and Szyma\'nski, M.~1998, {\it Acta Astron.},
     {\bf 48}, 269 (WS98).\par
\ref Wo\'zniak, P.R., Udalski, A., Szyma\'nski, M., Kubiak, M.,
     Pietrzy\'nski, G., Soszy\'nski, I., and \.Zebru\'n, K.~2001,
     {\it Acta Astron.}, {\bf 51}, 175.\par
\ref Wo\'zniak, P.R., Udalski, A., Szyma\'nski, M., Kubiak, M.,
     Pietrzy\'nski, G., Soszy\'nski, I., and \.Zebru\'n, K.~2002,
     {\it Acta Astron.}, {\bf 52}, 129.\par
\ref \.Zebru\'n, K.~1998, {\it Acta Astron.}, {\bf 48}, 289
     (Z98).\par

}

\end{document}